# Towards a Bayesian framework for option pricing

Henryk Gzyl*    Enrique ter Horst†    Samuel W. Malone‡

February 1, 2008

**Abstract**

In this paper, we describe a general method for constructing the posterior distribution of an option price. Our framework takes as inputs the prior distributions of the parameters of the stochastic process followed by the underlying, as well as the likelihood function implied by the observed price history for the underlying. Our work extends that of Karolyi (1993) and Darsinos and Satchell (2001), but with the crucial difference that the likelihood function we use for inference is that which is directly implied by the underlying, rather than imposed in an ad hoc manner via the introduction of a function representing "measurement error." As such, an important problem still relevant for our method is that of model risk, and we address this issue by describing how to perform a Bayesian averaging of parameter inferences based on the different models considered using our framework.

**JEL CLassification:** C11, C13, G12

**Keywords:** Bayesian analysis; Lévy processes; Option pricing; Risk-Neutral measure; Model selection; Bayesian nonparametrics.

*Henryk Gzyl, Instituto de Estudios Superiores de Administración IESA, Caracas, DF, Venezuela. henryk.gzyl@iesa.edu.ve

†Enrique ter Horst, Instituto de Estudios Superiores de Administración IESA, Caracas, DF, Venezuela. enrique.terhorst@iesa.edu.ve

‡Samuel W. Malone, Oxford University, Oxford, United Kingdom. samuel.malone@balliol.oxford.ac.uk

# Acknowledgments

The authors would like to thank German Molina and Robert L. Wolpert for helpful conversations.

# 1  Introduction

In this article, we present a novel method for "integrating out" parameters from the risk-neutral pricing formula of an option. We are not the first to propose a method that uses this well-known Bayesian technique for the problem of option pricing. Previous innovations in the area of Bayesian econometrics on techniques for integrating out parameters after the risk-neutral pricing formula for the option has been derived in closed form include Eraker et al (2000). However, no work has been done, to our knowledge, on integrating out the parameters during the transformation of the physical measure $\mathbb{P}$ to the risk neutral measure $\mathbb{Q}$, except when testing sequentially a precise hypothesis concerning the drift of a Brownian motion as in Rui (2002), or computing Bayes factors between different models as in Polson and Roberts (1994).

Karolyi (1993), Darsinos and Satchell (2001) are perhaps the previous work most closely related to our own. In their articles, the authors use the Bayesian technique of integrating out parameters to derive the posterior distribution in closed-form for a European call option when the underlying follows a geometric Brownian motion. To do this, however, they use the sufficiency property of the unbiased estimator of $\sigma^2$ from discretely sampled observations. The method we propose differs crucially in this respect: to obtain the posterior of the option price, we avoid what may be considered an inconsistency in previous methods, by using the likelihood implied by the stochastic process of the underlying. This is important on practical grounds, because the different likelihoods, except in a special case, will lead to different posterior distributions for the parameters, and this affects inference.

In general, the methodology developed in this paper is able to extend and yield the posterior distribution, in closed-form, as in the work of Karolyi (1993) and Darsinos and Satchell (2001). We construct these posterior distributions by combining the likelihood function that is implied by the underlying stochastic process with the prior distributions that are specified as the views of the market participant.

The method described in the remainder of our paper serves as an illustration of the linkages to be made between the broad areas of mathematical finance, on one hand, and Bayesian probability, on the other. We take a continuous time point of view for the purposes of exposition, but the technique can be formulated in discrete time and implemented for practical purposes. This is a subject of future research.

The outline of this paper is as follows. Section 2 reviews the literature on Bayesian option pricing and motivates our framework. Section 3 presents the methodology and describes how to find a likelihood function ($\mathbb{Q}$ the Radon-Nikodym derivative of the risk neutral measure with respect to the physical measure $\mathbb{P}$) by the use of the Esscher transform. It also addresses the question as how to choose acceptable prior distributions for continuous-time finance model parameters in order to perform a Bayesian analysis. Section 4 illustrates the methodology with two examples pertaining to the classical Black and Scholes model, as well as in a diffusion case. Section 4 concludes.

## 2 Previous work on Bayesian option pricing

The mainstream Bayesian literature has concerned itself with using state-space models as a way to get posterior distributions for derivatives perturbed around a Black & Scholes price of the following sort:

$$\log\left(\frac{S_t}{S_{t-1}}\right) = \mu + \sigma(W_t - W_{t-1}) \quad (1)$$

$$C_t = BS(\sigma, S_t) + \epsilon_t \quad (2)$$

where $W_t$ is a Brownian motion, the error term $\epsilon_t \sim N(0, \nu^2)$, and $BS(\sigma, S_t)$ is the option price from the classical Black and Scholes model. A more in-depth discussion of this approach can be found in pp. 35-36 of Johannes and Polson (2002). Both Johannes and Polson (2002), as well as Darsinos and Satchell (2001) obtain the posterior distribution of the volatility parameter $\sigma$ from the discrete time version of the continuous time process. Although this posterior distribution exists in discrete time, however, in continuous time it is a degenerate point mass, as Polson and Roberts (1994) explain, under a particular reference probability measure.

In order to get the posterior distribution of the theoretical Black and Scholes price, Johannes and Polson (2002) use a perturbation $\epsilon_t$ around the theoretical Black and Scholes price to construct a likelihood. Our framework consists of retrieving the likelihood function directly from the Radon-Nikodym process ($Z_t^{\theta^\star} \equiv \frac{d\mathbb{Q}}{d\mathbb{P}} |_{\mathcal{F}_t}$) used when performing a change of measure from to the physical measure $\mathbb{P}$ to the risk-neutral measure $\mathbb{Q}$, and where $\theta^\star$ is a function of the vector parameter $\theta$ governing the probability distribution of the logarithm of the exponential Lévy process $S_t$. Combining our likelihood $(Z_t^{\theta^\star})^{-1} \equiv \frac{d\mathbb{P}}{d\mathbb{Q}} |_{\mathcal{F}_t}$ with a prior $\pi(\theta)$ enables us to derive a posterior distribution for $\pi(\theta|\{S_s : 0 \leq s \leq t\})$ given an observed price history $\{S_s : 0 \leq s \leq t\}$. Darsinos and Satchell (2001), working within a Black and Scholes model and using the density of the distribution of the underlying with respect to the Lebesgue measure, are able to find a closed-form solution for the posterior of the call option. Our method can be carried out with the use of numerical simulations from the parameter's posterior distribution by standard Bayesian numerical methods as we illustrate throughout the paper.

Besides allowing us to integrate prior beliefs about the price process parameters together with the likelihood inherent in the price process to generate a marginal distribution for the price of the option, our method has another theoretical motivation as well. In the method summarized by equation 1 and equation 2, the parameters $\sigma$ and $\nu$ are estimated jointly using the call option price data and the price data for the underlying (see Johannes and Polson (2002) for the details). As a consequence, both of these parameters capture the joint effects of risk aversion and price volatility on option pricing, and it is somewhat unclear how to sort out these effects in that model. This is not surprising, since in that case the error term $\epsilon_t$ is interpreted as some sort of observation error.

The classical framework of option pricing supposes a call option $C(t, S_t, \theta)$ whose payoff $H(S_T)$ depends on our underlying $S_t$, and can be computed via the following integration [1]:

$$C(t, S_t, \theta) = \exp(-r(T-t)) E_\mathbb{Q}\{H(S_T) \mid \mathcal{F}_t\}$$

where integration is performed under the risk-neutral measure $\mathbb{Q}$, such that the discounted stock price $\exp(-rt) S_t$ is a $\mathbb{Q}$-martingale. General integration theory states that the following change of measure is also possible by invoking

---

[1] For a European call option, the payoff function is equal to $\max(S_T - K, 0)$ where $K$ is the strike price at termination date $T$.

the Radon-Nikodym theorem:

$$C(t, S_t, \theta) = \exp(-r(T-t)) \frac{E_\mathbb{P}\{Z_T^{\theta^\star} H(S_T) \mid \mathcal{F}_t\}}{E_\mathbb{P}\{Z_T^{\theta^\star} \mid \mathcal{F}_t\}}$$

and if considering a prior $\pi(d\theta)$, we can perform the following integration with respect to the prior $\pi(d\theta)$:

$$\begin{aligned}
C(t, S_t) &= \exp(-r(T-t)) \int_\Theta C(t, S_t, \theta) \pi(d\theta) \\
&= \exp(-r(T-t)) \int_\Theta \pi(d\theta) \frac{E_\mathbb{P}\{Z_T^{\theta^\star} H(S_T) \mid \mathcal{F}_t\}}{Z_t^{\theta^\star}} \\
&= \exp(-r(T-t)) \int_\Theta \pi(d\theta) (\frac{d\mathbb{P}}{d\mathbb{Q}} \mid_{\mathcal{F}_t}) E_\mathbb{P}\{Z_T^{\theta^\star} H(S_T) \mid \mathcal{F}_t\} \\
&= \exp(-r(T-t)) \int_\Theta g(S_u : u \leq t) \pi(d\theta \mid \mathcal{F}_t) E_\mathbb{P}\{Z_T^{\theta^\star} H(S_T) \mid \mathcal{F}_t\}
\end{aligned}$$

where $\theta \in \Theta$, $\Theta$ is the parameter space, $(Z_t^{\theta^\star})^{-1} = \frac{d\mathbb{P}}{d\mathbb{Q}} \mid_{\mathcal{F}_t}$, and

$$g(S_u : u \leq t) = \int_\Theta \pi(d\theta) \frac{d\mathbb{P}}{d\mathbb{Q}} \mid_{\mathcal{F}_t}$$

is the marginal probability distribution of the underlying $S_t$, which is thus a constant given the information $\mathcal{F}_t$.

The posterior distribution $\pi(d\theta \mid \mathcal{F}_t)$ comes from a direct application of Bayes rule:

$$\begin{aligned}
\pi(d\theta \mid \mathcal{F}_t) &= \frac{\pi(d\theta)(\frac{d\mathbb{P}}{d\mathbb{Q}} \mid_{\mathcal{F}_t})}{g(S_u : u \leq t)} \\
\pi(d\theta \mid \mathcal{F}_t) g(S_u : u \leq t) &= \pi(d\theta)(\frac{d\mathbb{P}}{d\mathbb{Q}} \mid_{\mathcal{F}_t})
\end{aligned}$$

This interesting result shows that in order to integrate out the uncertainty related to the governing parameters from the probability distribution of the underlying, one needs to use the likelihood that comes automatically by the specification of underlying through $\frac{d\mathbb{P}}{d\mathbb{Q}} \mid_{\mathcal{F}_t}$.

It is often the case that competing models could have generated the underlying process $S_t$. Let us suppose that we have $k$ of those competing models[2] each of which has a prior probability of $\mathbb{P}(M_i)$, a Radon-Nikodym derivative $\frac{d\mathbb{P}^i}{d\mathbb{Q}^i}|_{\mathcal{F}_t}$, and a model parameter vector $\theta^i$, for $i = 1, \cdots, k$. These vectors $\theta^i$ might have different meanings depending on the model they refer to. If we wish to see which of the models models the data best, then one could compute their posterior probabilities. The marginal likelihood $\mathbb{P}(S_t \mid M_i)$ of model $M_i$ gives a measure of observing the data given that $M_i$ is true. Each different competing model will give rise to a different vector parameter $\theta^i$, having similar or completely different interpretations. This last quantity can be computed as follows:

$$\mathbb{P}(S_u : u \leq t \mid M_i) = \int_\Theta (\frac{d\mathbb{P}^i}{d\mathbb{Q}^i}|_{\mathcal{F}_t}) \pi(d\theta^i).$$

---

[2] Example of such models could be the case where the underlying process $S_t$ is described either by a jump-diffusion, diffusion, or by a pure jump Lévy process, or by the same process with different parameters.

The marginal likelihood together with the prior probability on model $i$ enables to use Bayes theorem which updates our prior beliefs that the data comes from model $M_i$ as follows:

$$\mathbb{P}(M_i \mid S_u : u \leq t) = \frac{\mathbb{P}(S_u : u \leq t \mid M_i)\mathbb{P}(M_i)}{\sum_{i=1}^{k} \mathbb{P}(S_u : u \leq t \mid M_i)\mathbb{P}(M_i)}$$

As Cont (2006) points out, model uncertainty regarding option pricing models leads to "model risk" which can be seen as an extension of model misspecification. Bayesian model averaging is a way to incorporate model uncertainty for option prices, where one could compute an option price integrating model uncertainty in the following way:

$$\tilde{C}(t, S_t) = \sum_{i=1}^{k} C(t, S_t \mid M_i)\mathbb{P}(M_i \mid S_u : u \leq t)$$

where the $C(t, S_t \mid M_i)$ are the option prices $C(t, S_t)$ computed under model $M_i$. $\tilde{C}(t, S_t)$ would thus yield a model weighted "average" option price.

Although we do not always have a closed-form solution for the integral $\int_\Theta (\frac{d\mathbb{P}^i}{d\mathbb{Q}^i}|_{\mathcal{F}_t})\pi(d\theta^i)$, we can approximate it through classical Markov Chain Monte Carlo methods to get the marginal likelihood. In order to do this, we need a likelihood and a prior distribution on the model parameters in order to perform a Bayesian analysis. In the next section, we show how to find this likelihood.

# 3 The Framework

## 3.1 Preview of coming attractions

As motivation for the material in the following sections, we present a version of the Black-Scholes model in discrete time, which stands between the binomial model and the continuous time Black-Scholes model, in which the posteriorization procedure does not involve stochastic calculus. Assume that our price process is described by a sequence $S_n = e^{\mu+R_n} S_{n-1}$, and $S_0 \in \mathbb{R}$ is given, and where the sequence $\{R_n \,|\, n \geq 1\}$ is an $i.i.d$ sequence of $N(0,1)$ random variables. We take $(\Omega = \mathbb{R}^{\mathbb{N}}, \mathcal{F}_n, \mathcal{F}, P)$ as sample space, where $R_n$ will denote the coordinate mappings and $\mathcal{F}_n = \sigma\{R_k \,|\, k \leq n\}$, $\mathcal{F} = \sigma\{R_k \,|\, k < \infty\}$ and $P$ is the obvious infinite product of gaussian measures on $(\Omega, \mathcal{F})$.

It is a simple calculation to verify that the value of $a$ such that

$$E[e^{aR-a^2\sigma^2/2} e^{\mu+R_n}] = e^r$$

is $a = \left(\frac{r-\nu}{\sigma^2}\right)$ where $\nu = \mu - \sigma^2/2$. Also, it is a standard exercise to verify that the limit $Z$ of the Wald martingale $Z_n \equiv \prod_{i=1}^{n} e^{aR_n - a^2\sigma^2/2}$ provides us with a measure $Q \sim P$ such that $Z_n = dQ/dP|_{\mathcal{F}_n}$ and that the present value price process $S_n^* = S_n e^{-r}$ is a $Q$-martingale.

It is also simple to verify that the likelihood $Z_n^{-1} = e^{-a\sum_1^n R_k + na^2\sigma^2/2} = dP/dQ|_{\mathcal{F}_n}$ is a $Q$-martingale. The interest in $Z_n^{-1}$ is that it can be thought of as the conditional density $\frac{dP}{dQ}(S_0, ..., S_n \,|\, \mu)$. To see how this comes about, note that $\sum_{k=1}^{n} R_k = \sum_{k=1}^{n} \ln\left(S_k/S_{k-1}\right) - n\mu$, and after completing squares and re-arranging the exponent we obtain

$$\begin{aligned}
Z_n^{-1}(S_0, ..., S_n \,|\, \mu) &= \exp\Big[\frac{n}{2\sigma^2}\big(r + \frac{\sigma^2}{2} - \ln\left(S_n/S_0\right)/n\big)^2\Big] \times \\
&\quad \times \exp\Big[-\frac{n}{2\sigma^2}\big(\nu + \frac{\sigma^2}{2} - \ln\left(S_n/S_0\right)/n\big)^2\Big].
\end{aligned}$$

From this point on, the marginalization to obtain the posterior of $\mu$ given $S_0, ..., S_n$ is a routine matter in the gaussian set up. Let us now proceed to the continuous time case.

## 3.2 Change of measures and the likelihood function in option pricing

It is common to use the following form when modelling the underlying $S_t$ of a derivative:

$$S_t = \exp(X_t)$$

where $X_t$ can either be a Lévy process or a diffusion, and thus $S_t$ is the exponential[3] of this process discounted by $\exp(-rt)$. In order to price options, one needs to find an equivalent probability measure $\mathbb{Q} \sim \mathbb{P}$ such that $S_t \equiv \exp(-rt + X_t)$ is a martingale under $\mathbb{Q}$.

When performing a change of measure for a given stochastic process $S_t$ under $\mathbb{P}$ to $\mathbb{Q}$, one can regard $S$ as a random variable on the space $\Omega = \mathbb{D}[0, +\infty)$ of càdlàg (Continue à droite limites à gauche) paths together with its associated filtration $(\mathcal{F}_t)_{t \geq 0}$. This measure $\mathbb{P}$ is therefore defined on the space of sample paths of $X$, and so the Radon-Nikodym derivative $\frac{d\mathbb{P}}{d\mathbb{Q}} |_{\mathcal{F}_t}$ with respect to the reference measure $\mathbb{Q}$ is the likelihood function [4] after the process has been observed up to time $t$.

It is easy to verify that the independence of the increments of $X$ imply that $Z_t^{\theta_\star} = \exp(\theta_\star X_t - tk(\theta_\star))$. Furthermore, $Z_t^{\theta_\star}$ is not only a positive $\mathcal{F}_t$-martingale under $\mathbb{P}^\theta$, but happens to be the likelihood function given time $t$. What is interesting here is that $S_t^* \equiv S_t / \exp(rt) = \exp(X_t - rt)$ is a $Q^{\theta_\star}$-martingale for every $\theta$ as is easy to verify. This observation will enable us to use Bayes theorem in the following sections to compute posterior distributions of the parameters that govern the dynamics of the underlying $S_t$. Since we will mostly work with continuous stochastic processes, we set $\Omega = \mathcal{C}[0, \infty)$, and[5] $(\Omega, \mathcal{F}, \mathbb{P})$ is the standard Wiener space.

---

[3]See Applebaum (2004) and Oksendal (2003) for the case when $X_t$ is a Lévy process and a diffusion respectively.

[4]See chapter X, section 2 of Jacod and Shiryaev (1987) on the equivalence between Radon-Nikodym derivatives and likelihood function.

[5]$\mathcal{C}[0, \infty)$ is the space of continuous functions defined from $[0, \infty)$ into $\mathbb{R}$.

# 4 Parameter posteriors for some popular continuous time models of the underlying

## 4.1 Geometric Brownian motion

In the classic paper by Black and Scholes (1973), the stock price $S_t$ is solution to the following SDE:

$$\frac{dS_t}{S_t} = \mu dt + \sigma dW_t \tag{3}$$

whose solution is equal to:

$$S_t = S_0 \exp\left[(\mu - \frac{\sigma^2}{2})t + \sigma W_t\right] \tag{4}$$

Working with the discounted stock price, we obtain

$$\exp(-rt) S_t = S_0 \exp\left[(\mu - r - \frac{\sigma^2}{2})t + \sigma W_t\right] \tag{5}$$

Invoking the Cameron-Martin-Girsanov (CMG) theorem enables us to determine the deterministic risk-neutral condition $\mu = r$ which is the one that defines the unique martingale measure. We then get that $S_t$ is a martingale[6] under $\mathbb{Q}$ and is equal to:

$$S_t = S_0 \exp\left[(r - \frac{\sigma^2}{2})t + \sigma W_t\right] \tag{6}$$

As it turns out, the martingale condition $\mu = r$ holds asymptotically in our Bayesian framework, as t tends to infinity, but for all finite $t$ the value of $\mu$ consistent with no-arbitrage has a posterior distribution that is non-degenerate.

**Theorem 1.** *Under the model given by equation 3, from the Cameron-Martin-Girsanov theorem the density process $Z_t^\theta$ is given by*

$$Z_t^\theta = \exp\left[W_t\left(\frac{r-\mu}{\sigma}\right) - \frac{t}{2}\left(\frac{r-\mu}{\sigma}\right)^2\right] \tag{7}$$

---

[6]$S_t$ is a $\mathbb{Q}$-martingale, which is equivalent to showing that $Z_t^\theta S_t$ is a $\mathbb{P}$-martingale.

and the posterior distribution for the drift parameter $\mu$ is given by

$$[\mu|\{S_s : 0 \leq s \leq t\}] \sim N\left[r - \frac{\sigma W_t}{t}, \frac{\sigma^2}{t}\right] \tag{8}$$

*Proof.* See Appendix. □

Theorem (1) illustrates that, even in the setting where markets are complete and we can sample continuously, we obtain a posterior distribution for the drift parameter $\mu$ that differs from the usual no-arbitrage condition that $\mu = r$ with probability 1. We recover this no-arbitrage condition in the limit as time goes to infinity, and when time is finite, we have that if $W_t > 0$, then $E[\mu|\{S_s : 0 \leq s \leq t\}] = r - \frac{\sigma W_t}{t} < r$, and vice versa.

One important observation is that as $t \to +\infty$ we get $\frac{\sigma^2}{t} \to 0$ and thus $\mu \to \theta_0 = r$ since $\frac{\sigma W_t}{t} \to 0$ as $t \to +\infty$. Therefore, the posterior distribution $[\mu|\{S_s : 0 \leq s \leq t\}]$ is consistent at $r = \theta_0$. This is a property that a posterior distribution should have in general if we wish to learn more about a parameter, and is the Bayesian analogue of consistency for an estimator in classical statistics. We now do a variation on the definition of consistency from Ghosh and Ramamoorthi (2003) and propose the following one:

**Definition 1.** *For each $t$, let $\pi(\theta|S_s : 0 \leq s \leq t)$ be a posterior probability distribution given $\{S_s : 0 \leq s \leq t\}$. The collection $\{\pi(\theta|S_s : 0 \leq s \leq t)\}$ is said to be consistent at $\theta_0$ if there is a $\Omega_0 \subset \Omega = \mathcal{C}[0, \infty)$ with $\mathbb{P}_{\theta_0}(\Omega_0) = 1$ such that if $\omega$ is in $\Omega_0$, then for every neighborhood $U$ of $\theta_0$,*

$$\pi(U|S_s : 0 \leq s \leq t) \to 1 \quad a.s. \ \mathbb{P}_{\theta_0} \tag{9}$$

where $\mathbb{P}_{\theta_0}$ is a probability measure defined on the space of right continuous functions with left limits, and $S_t$ is the underlying process. When performing a Bayesian analysis, it is important for the posterior to converge to a degenerate point mass at the unknown parameter $\theta_0$, which means in our setup, the posterior variance tends to 0. In the previous example, we obtain consistency as the posterior variance $\frac{\sigma^2}{t} \to 0$. It is worth to notice that $Z_t^\theta$ depends on the initial and final values $S_0$ and $S_0$ through $\log\left(\frac{S_t}{S_0}\right)$. Working with continuously compounded returns enables us to use the same definition of consistentcy as in Ghosh and Ramamoorthi (2003), as well as to prove that two agents having different prior distributions $\pi(\theta)$ reach in the end the same conclusions regarding $\theta$. Throughout our examples we shall show that our posterior distributions are consistent for both $\mu$ and $\sigma^2$. We now cite from Ghosh and Ramamoorthi (2003) the following second definition of consistency as well as a theorem:

**Definition 2.** *For each $n$, let $\pi(\theta|R_t^1, \ldots, R_t^n)$ be a posterior given the iid sequence $R_t^1, \ldots, R_t^n$. The sequence $\{\pi(\theta|R_t^1, \ldots, R_t^n)\}$ is said to be consistent at $\theta_0$ if there is a $\mathbb{W}_0 \subset \mathbb{W}$ with[7] $\mathbb{P}_{\theta_0}^\infty(\Omega_0) = 1$ such that if $\omega$ is in $\Omega_0$, then for every neighborhood $U$ of $\theta_0$,*

$$\pi(U|R_t^1, \ldots, R_t^n) \to 1 \ \ a.s. \ \ \mathbb{P}_{\theta_0}^\infty \tag{10}$$

where $R_t^i \equiv \log(\frac{S_t^i}{S_0^i})$ is the return during the $i$-th interval of length $t$. $S_t^i$ and $S_0^i$ are the prices at the beginning and the end of the $i$-th interval of length $t$. We have $n$ of those independent returns since for a given time interval $[0, T]$ such that $T = nt$, we use underlying processes for $S_t$ with Stationary and Independent Increments.

What if two different agents used the same probability model that generated the underlying price process $S_t$, but had different prior distributions regarding the parameters governing the latter. Would both posterior distributions be consistent at the same value $\theta_0$? In other words, under what conditions would two different priors lead to the same inference? In order to answer the latter, we cite from Ghosh and Ramamoorthi (2003) the general definition of consistency for a posterior distribution, together with a theorem.

**Theorem 2.** *Assume that the family $\{\mathbb{P}_t^\theta : \theta \in \mathbb{A}\}$ is dominated by a $\sigma$-finite measure $\mu$ and let $p_\theta$ denote the density of $\mathbb{P}_t^\theta$. Let $\theta_0$ be an interior point of $\Theta$, and $\pi_1, \pi_2$ be two priors densities with respect to a measure $\nu$, which are positive and continuous at $\theta_0$. Let $\pi(\theta|R_t^1, \ldots, R_t^n)_i$, $i = 1, 2$ denote the posterior densities of $\theta$ given $\{R_t^1, \ldots, R_t^n\}$. If $\pi(\theta|R_t^1, \ldots, R_t^n)_i$, $i = 1, 2$ are both consistent at $\theta_0$ then:*

$$\lim_n \int |\pi_1(\theta|R_t^1, \ldots, R_t^n) - \pi_2(\theta|R_t^1, \ldots, R_t^n)|d\nu(\theta) = 0 \ \ a.s. \ \ \mathbb{P}_{\theta_0} \tag{11}$$

where $\mathbb{P}_t^\theta$ is the probability distribution of the return $R_t = \log(\frac{S_t}{S_0})$.

---

[7]Here $\mathbb{W} = \mathbb{R}^\infty$ is the space of infinite sequences, $\mathbb{P}_{\theta_0}^\infty = \otimes \mathbb{P}_{\theta_0}$ is the product probability measure defined on the sigma algebra $\mathcal{B}(\mathbb{R}^\infty)$ of $\mathbb{W}$. Here $\mathbb{P}_{\theta_0}$ is the probability distribution of the returns $R_t^i$ with respect to $\mathbb{Q}_{\theta_0}$.

This last theorem shows that as $n$ increases (and therefore our information set increases), the importance of the prior distribution fades away (and thus the importance of the prior's hyperparameters as well), since we get to observe more and more data. In our framework, the dominating measure $\mu$ of Ghosh and Ramamoorthi (2003) is our $\mathbb{Q}$ and their family $\mathbb{P}_t^\theta$ corresponds to our physical probability measure regarding the returns $R_t$.

## 4.2 A Bayesian Inference of the Black & Scholes model

As we just saw in the previous section, the log returns of the stock price $\frac{\log(\frac{S_t}{S_0})}{t} \sim N\left[\mu - \frac{\sigma^2}{2}, \frac{\sigma^2}{t}\right]$ are normally distributed[8], thus the conditional likelihoods for $\mu$ and $\sigma^2$ in the Black & Scholes model are proportional to (see appendix):

$$L(\sigma^2 \mid \log(\frac{S_t}{S_0}), \mu) \propto (\sigma^2)^{(\frac{1}{2}-1)} \exp\left\{-\frac{1}{2}\left[\frac{t(\mu - \frac{\log \frac{S_t}{S_0}}{t})^2}{\sigma^2} + \frac{t\sigma^2}{4}\right]\right\} \quad (12)$$

$$L(\mu \mid \log(\frac{S_t}{S_0}), \sigma^2) \propto \exp\left\{-\frac{t}{2\sigma^2}\left[\frac{\log \frac{S_t}{S_0}}{t} - (\mu - \frac{\sigma^2}{2})\right]^2\right\} \quad (13)$$

The first conditional likelihood given $\mu$ is nothing but proportional to a Generalized Inverse Gaussian with parameters $\lambda$, $\delta$, and $\gamma$ (in short $GIG(\lambda, \delta, \gamma)$) whose density is equal to:

$$f(x \mid \lambda, \delta, \gamma) = (\frac{\gamma}{\delta})^\lambda \frac{1}{2K_\lambda(\gamma\delta)} x^{(\lambda-1)} \exp\left\{-\frac{1}{2}\left[\frac{\delta^2}{x} + \gamma^2 x\right]\right\} \quad (14)$$

where we use the same parametrization as in Silva et al (2006). We note that equation (12) is proportional to a $GIG(\lambda, \delta, \gamma)$ with parameters equal to $\lambda = \frac{1}{2}$, $\delta^2 = t(\mu - \frac{\log \frac{S_t}{S_0}}{t})^2$, and $\gamma^2 = \frac{t}{4}$, and that the conditional posterior distribution of $\mu$ given $\sigma^2$ is normally distributed as $N\left[\frac{\log \frac{S_t}{S_0}}{t} + \frac{\sigma^2}{2}, \frac{\sigma^2}{t}\right]$. We prove these last statements and more in the following two lemmas whose proofs are in the appendix.

---

[8]Here our Radon-Nikodym derivative $\frac{d\mathbb{P}}{d\lambda}\mid_{\mathcal{F}_t}$ is with respect to Lebesgue measure $\lambda$. One can compute this ratio with respect to the probability measure $\mathbb{P}$ using the following chain rule: $\frac{d\mathbb{Q}}{d\mathbb{P}}\mid_{\mathcal{F}_t} = \frac{d\mathbb{Q}}{d\lambda}\mid_{\mathcal{F}_t} \frac{d\lambda}{d\mathbb{P}}\mid_{\mathcal{F}_t}$.

**Lemma 1.** *When choosing flat uniform prior distributions on $\mathbb{R}$ and $\mathbb{R}_+$ for $\mu$ and $\sigma^2$ in the Black and Scholes model, then their posterior conditional distributions are $\pi(\mu \mid \sigma^2, \log \frac{S_t}{S_0}) = N\left[\frac{\log \frac{S_t}{S_0}}{t} + \frac{\sigma^2}{2}, \frac{\sigma^2}{t}\right]$ and $\pi(\sigma^2 \mid \mu, \log \frac{S_t}{S_0}) = GIG(\lambda, \delta, \gamma)$ respectively, where $\lambda = \frac{1}{2}$, $\delta^2 = t(\mu - \frac{\log \frac{S_t}{S_0}}{t})^2$, and $\gamma^2 = \frac{t}{4}$, for uniform priors on defined on $\mathbb{R}$ and $\mathbb{R}_+$ respectively.*

*Proof.* See appendix. □

**Lemma 2.** *When chosing normal prior distributions ($\pi(\mu) = N[m, s]$) on $\mathbb{R}$ for $\mu$, and a $\pi(\sigma^2) = GIG(\lambda, \delta, \gamma)$ on $\mathbb{R}_+$ for $\sigma^2$ in the Black & Scholes model, their posterior conditional distributions are $\pi(\mu \mid \sigma^2, \log \frac{S_t}{S_0}) = N\left[\left(\frac{\log \frac{S_t}{S_0} + \frac{\sigma^2 t}{2} + \frac{m\sigma^2}{s^2}}{t + \frac{\sigma^2}{s^2}}\right), \frac{\sigma^2}{t + \frac{\sigma^2}{s^2}}\right]$ and $\pi(\sigma^2 \mid \mu, \log \frac{S_t}{S_0}) = GIG(\lambda', \delta', \gamma')$, where $\lambda' = \lambda - \frac{1}{2}$,*
*$\delta'^2 = t(\mu - \frac{\log \frac{S_t}{S_0}}{t})^2 + \delta^2$, and $\gamma'^2 = \frac{t}{4} + \gamma^2$.*

*Proof.* See appendix. □

When modelling the underlying with a given process in continuous time (in this example a Geometric Brownian Motion), there is already a likelihood function implied by this latter. This methodology proposes an extension of Karolyi (1993), Darsinos and Satchell (2001), where their posterior for $\sigma^2$ is an Inverse Gamma distribution, which appears to be a specific case of the $GIG(\lambda, \delta, \gamma)$ as explained in Silva et al (2006). Furthermore, both Karolyi (1993), Darsinos and Satchell (2001) construct the likelihood for $\sigma^2$ relying on the mathematical result that when log returns are normally distributed, the statistic[9] $\frac{\nu s^2}{\sigma^2} \sim \chi^2(\nu)$ and independent of $\mu$ since the latter is sufficient. Our extension relies on a more general mathematical framework that consists of working with the Radon-Nikodym derivative $Z_t$ instead[10].

With this example we illustrate one of the core points of this paper that one is not free to elicit a likelihood when imposing a mathematical model for the underlying stochastic process, which already has one. These two likelihoods can lead to different inferences for the parameter $\sigma^2$ unless they are by coincidence proportional one to another as Berger and Wolpert (1984) point out.

---

[9]$\chi^2(\nu)$ is a chi-square random variable with $\nu = n - 1$ degrees of freedom and $s^2 \equiv \frac{1}{n-1}\sum_{i=1}^{n}(R_t^i - \bar{R}_t)^2$.

[10]Here the reference measure is just Lebesgue measure $\lambda$.

From the Cameron-Martin-Girsanov theorem, the likelihood function for a general diffusion process $dX_t = f(\theta, t, X_t)dt + \sigma(t, X_t)dW_t$ is given by[11]:

$$\begin{aligned}
Z_t &= \exp\left\{\int_0^t \frac{(r - f(\theta, s, X_s))}{\sigma(s, X_s)} d\log\left(\frac{S_t}{S_0}\right)\right\} \times \\
&\quad \times \exp\left\{-\frac{1}{2}\int_0^t \frac{(r - f(\theta, s, X_s))(r + f(\theta, s, X_s) - \sigma^2)}{\sigma^2} ds\right\}
\end{aligned}$$

where $S_t = S_0 \exp\left\{X_t - \frac{1}{2}\int_0^t \sigma(s, X_s)^2 ds\right\}$ is the stock price process. Polson and Roberts (1994) compute posterior probability distributions as well as Bayes factors for the drift of the diffusion when $f(\theta, t, X_t) = \theta f(t, X_t)$. Our posterior distribution for the drift reduces to example 4 in Polson and Roberts (1994) when $r = 0$. We not only develop a Bayesian methodology for consistently estimating parameters in the context of option pricing, but also compute the posterior probability distributions for both $\sigma^2$ and $\mu$, as well as in continuous time, extending the work by Karolyi (1994).

## 5 Discussion and Conclusions

When traders and market participants use pricing formulas for derivatives, the price they obtain is a function of the parameter values that they assume. However, it is reasonable to expect that these parameter values are not known with certainty. As a result, it is worthwhile to take this uncertainty about parameter values into account in option pricing. The natural way of doing this is through the use of Bayesian methods. In this paper, we present a framework for Bayesian option pricing that can be applied to a very general set of stochastic processes for the underlying. Using directly the probability model for the stochastic process of the stock price process, Girsanov's theorem helps us derive the likelihood function which one of the key ingredients that enables us to yield posterior probability distributions for model parameters.

We show that in our framework, in which the posterior distribution for the parameter estimates is computed using the likelihood implied by the underlying stochastic process and prior distribution that represents agents´ beliefs about the parameter values, the likelihood function is constructed from a martingale. Asymptotically, as time goes to infinity and we collect more data, our prior beliefs can be updated. Moreover, we discuss our framework with the example of geometric Brownian motion, where our findings are consistent with those of Polson and Roberts (1994) in the context of the posterior distribution of the drift $\mu$. We extend the methodology by Karolyi (1993), by noticing how to retrieve the implied likelihood imposed by the stochastic process of the underlying. In this sense we show that when working with derivatives, it is important to have coherence between the likelihood implied by the stochastic process of the underlying and the likelihood of the econometric model used for inference.

---

[11]See Oksendal (2003).

In terms of future research, the most worthwhile application of these techniques would probably be to the case of a tractable version of a general Lévy process, which in continuous time yields nontrivial likelihoods for the case of jump-diffusion processes for all of the parameters, as we see in the appendix. In addition, a potentially fruitful application of this technique in discrete time, to binomial tree pricing methods, is the subject of current research.

# Appendix

### Posterior for $\mu$

It is a standard computation to show that the value of $a \in \mathbb{R}$ such that $Z_t^a \equiv \exp\left(aW_t - \frac{a^2 t}{2}\right)$ is a density for the risk neutral measure $\mathbb{Q}$ making $Z_t^a S_t \exp(-rt)$ into a $\mathbb{P}$-martingale is $a = \frac{r-\mu}{\sigma}$. If $\theta = (\mu, \sigma^2)$ are the model parameters, then $\theta_\star = a(\theta) = \frac{r-\mu}{\sigma^2}$.

We see how $\theta_\star$ is a function of $\theta \equiv (\mu, \sigma)$ and makes $S_t = \exp(X_t)$ into a $\mathbb{Q}$-martingale, where the Radon-Nikodym derivative $\frac{d\mathbb{Q}}{d\mathbb{P}}\mid_{\mathcal{F}_t}$ is equal to:

$$\frac{d\mathbb{Q}}{d\mathbb{P}}\mid_{\mathcal{F}_t} = \exp\left(\theta_\star X_t - \left[\theta_\star(\mu - \frac{\sigma^2}{2}) + \frac{\sigma^2 \theta_\star^2}{2}\right]t\right).$$

Substituting $\theta_\star$ in the Radon-Nikodym derivative and using a flat uniform prior for $\mu$, the posterior is proportional to:

$$\frac{d\mathbb{Q}}{d\mathbb{P}}|_{\mathcal{F}_t} \propto \exp\left(-\frac{t}{2\sigma^2}[\mu^2 - 2\mu(r - \frac{\sigma W_t}{t})]\right)$$

$$\propto \exp\left(-\frac{t}{2\sigma^2}[\mu - (r - \frac{\sigma W_t}{t})]^2\right).$$

We conclude that the posterior distribution for $\mu$ is:
$N\left[r - \frac{\sigma W_t}{t}, \frac{\sigma^2}{t}\right]$

## A  Posterior probability distributions for Jump-Diffusion processes

Consider first the case where the price process is $S_t = e^{\nu t + \sigma W_t}$, with $W$ as above and $\nu = \mu - \sigma^2/2$. In this case the getting $(Z_t^{\theta_\star})^{-1}$ ready for posteriorization will be pretty much like in section 3.1 Note to begin with that from $dS_t = \mu S_t dt + \sigma S_t dW_t$ and Îto's formula that on the one hand $\sigma dW_t = dS_t/S_t - \mu dt$ and on the other hand, therefore

$$d(\ln\left(S_t/S_0\right) + \frac{\sigma^2 t}{2}) = dS_t/S_t$$

and therefore that
$$\sigma W_t = \ln\left(S_t/S_0\right) - t\nu.$$

With all this, just some arithmetics allows us to rearrange the exponent in $(Z_t^{\theta_\star})^{-1} = \exp\left[-\frac{\theta_\star}{\sigma}\sigma W_t + \frac{(\theta_\star)^2 \sigma^2 t}{2}\right]$, where recall, $\theta_\star = (r-\mu)/\sigma$, so that

$$(Z_t^{\theta_\star})^{-1} = \exp\left[\frac{t}{2\sigma^2}(r - \frac{\sigma^2}{2} + \frac{\ln\left(S_t/S_0\right)}{t})^2 - \frac{t}{2\sigma^2}(\mu - \frac{\sigma^2}{2} + \frac{\ln\left(S_t/S_0\right)}{t})^2\right].$$

When the asset prices are driven by discontinuous factors, part of the routine is easy, but part is not. Assume for example that $dS_t = \mu S_t dt + \sigma S_t dW_t + dJ(t)$, where the first two terms in the right hand side are as in the previous example, whereas $J(t) = \sum_{n=1}^{N(t)} \xi_n$ is a compound Poisson process such that $N(t)$ is a Poisson Process with intensity $\lambda$ and the i.i.d. sequence is independent of both $W$ and $N$, and for the sake of simplicity, let us assume the $\xi_n$ to be bounded.

The price equation has solution

$$S_t = e^{t(\mu-\sigma^2/2)+\sigma W_t} \prod_{n=1}^{N(t)}(1+\xi_n) = e^{t(\mu-\sigma^2/2)+\sigma W_t + Y(t)}$$

where $Y_t = \sum_{n=1}^{N(t)} \ln(1+\xi_n) \equiv \sum_{n=1}^{N(t)} \eta_n$. Clearly we must assume that $(1+\xi_n) > 0$ for the price process to be positive. The boundedness assumption on the jumps yields the existence of $\int (e^{\theta x} - 1)n(dx)$, with $n(dx) = dP(\eta_1 \leq x)$. With this, clearly $E[e^{\theta(\sigma W_t + Y_t)}] = e^{tk(\theta)}$, with

$$k(\theta) = \frac{\sigma^2}{2} + \lambda \int (e^{\theta x} - 1)n(dx)$$

This time when we ask for the existence of a value $\theta^\star$ such that

$$E[e^{\theta(\sigma W_t + Y_t) - tk(\theta)} S_t] = e^{tr}$$

we are led to the equation $k(\theta_\star + 1) - k(\theta_\star) = r - \nu$, with $\nu$ as above. This amounts to solving

$$\theta_\star + \lambda \int (e^x - 1)e^{x\theta_\star} n(dx) = \frac{r - \nu}{\sigma^2}.$$

When $\theta_*$ ranges from $-\infty$ to $+\infty$, the left hand side of the identity above , ranges in an increasing way over the same range, thus the equation has a solution. So there is at least one risk neutral measure for the given asset price. Furthermore, it is not hard to see using Îto's formula that this time we also have

$$\sigma W_t + Y_t = \ln(S_t/S_0) - t\nu$$

but the likelihood

$$Z_t^{\theta_\star} = \exp^{-\theta_\star(\sigma W_t + Y_t) + tk(\theta_\star)} = e^{-\theta_\star\left(\ln(S_t/S_0) - t\nu\right) + tk(\theta_\star)}$$

does not lead to a posterior distribution for $\theta$ easy to sample from.

## B Proof of Lemmas 1 and 2

In this section we derive the full conditional posterior distributions for both $\mu$ and $\sigma^2$ in the Black & Scholes model. In their model, the log returns are normally distributed as $\frac{\log(\frac{S_t}{S_0})}{t} \sim N\left[\mu - \frac{\sigma^2}{2}, \frac{\sigma^2}{t}\right]$. Once we observe the whole sample path $S_t$, we get the following function for $\mu$ and $\sigma^2$:

$$L(\mu, \sigma^2 \mid \log(\frac{S_t}{S_0})) = \frac{1}{\sqrt{(2\pi\frac{\sigma^2}{t})}} \exp\left\{-\frac{t}{2\sigma^2}\left[\frac{\log \frac{S_t}{S_0}}{t} - (\mu - \frac{\sigma^2}{2})\right]^2\right\} \quad (15)$$

Proof for Lemma (1):

*Proof.* Since the priors for both $\mu$ and $\sigma^2$ are both flat priors, the joint posterior distribution is proportional to the likelihood function, which is equal to equation (15). Keeping only the terms in $\sigma^2$ in equation (15), we conclude easily that the conditional posterior distribution for $\sigma^2$ is a $GIG(\lambda', \delta', \gamma')$ where $\lambda' = \frac{1}{2}$, $\delta'^2 = t(\mu - \frac{\log \frac{S_t}{S_0}}{t})^2$, and $\gamma'^2 = \frac{t}{4}$. The conditional posterior distribution for $\mu$ is a $N\left[\frac{\log \frac{S_t}{S_0}}{t} + \frac{\sigma^2}{2}, \frac{\sigma^2}{t}\right]$. □

Proof for Lemma (2):

*Proof.* By Bayes rule, the posterior $\pi(\sigma^2 \mid \log(\frac{S_t}{S_0}), \mu)$ for $\sigma^2$ given $\mu$ and the data is proportional to:

$$\propto L(\sigma^2 \mid \log(\frac{S_t}{S_0}), \mu) \pi(\sigma^2)$$

$$\propto (\sigma^2)^{(\frac{1}{2}-1)} \exp\left\{-\frac{1}{2}\left[\frac{t\sigma^2}{4} + \frac{t(\mu - \frac{\log \frac{S_t}{S_0}}{t})^2}{\sigma^2}\right]\right\} \pi(\sigma^2)$$

and the likelihood $\pi(\sigma^2 \mid \mu, \log(\frac{S_t}{S_0}))$ times the $GIG(\lambda, \delta, \gamma)$ prior distribution $\pi(\sigma^2)$ yields:

$$(\sigma^2)^{(\lambda-\frac{1}{2}-1)} \exp\left\{-\frac{1}{2}\left[\sigma^2(\frac{t}{4} + \gamma^2) + \frac{(t(\mu - \frac{\log \frac{S_t}{S_0}}{t})^2 + \delta^2)}{\sigma^2}\right]\right\}$$

which is proportional to a $GIG(\lambda', \delta', \gamma')$, where $\lambda' = \lambda - \frac{1}{2}$, $\delta'^2 = t(\mu - \frac{\log \frac{S_t}{S_0}}{t})^2 + \delta^2$, and $\gamma'^2 = \frac{t}{4} + \gamma^2$.

The full conditional posterior distribution for $\mu$ given $\sigma^2$ is given by $\pi(\mu \mid \sigma^2, \log(\frac{S_t}{S_0}))$ and is proportional to:

$$\propto \pi(\mu) \exp\left\{-\frac{t}{2\sigma^2}\left[\frac{\log \frac{S_t}{S_0}}{t} - (\mu - \frac{\sigma^2}{2})\right]^2\right\}$$

$$\propto \exp\left\{-\frac{t}{2\sigma^2}\left[\frac{\log \frac{S_t}{S_0}}{t} - (\mu - \frac{\sigma^2}{2})\right]^2 - \frac{1}{2s^2}(\mu - m)^2\right\}$$

$$\propto \exp\left\{-\frac{(t + \frac{\sigma^2}{s^2})}{2\sigma^2}\left[\mu^2 - 2\mu\left(\frac{m\sigma^2 + ts^2\left(\frac{\log \frac{S_t}{S_0}}{t} + \frac{\sigma^2}{2}\right)}{ts^2 + \sigma^2}\right)\right]\right\}$$

$$\propto \exp\left\{-\frac{(t + \frac{\sigma^2}{s^2})}{2\sigma^2}\left[\mu^2 - 2\mu\left(\frac{m\frac{\sigma^2}{s^2} + \log \frac{S_t}{S_0} + \frac{t\sigma^2}{2}}{t + \frac{\sigma^2}{s^2}}\right)\right]\right\}$$

$$\propto \exp\left\{-\frac{(t + \frac{\sigma^2}{s^2})}{2\sigma^2}\left[\mu - \left(\frac{m\frac{\sigma^2}{s^2} + \log \frac{S_t}{S_0} + \frac{t\sigma^2}{2}}{t + \frac{\sigma^2}{s^2}}\right)\right]^2\right\}$$

where $\pi(\mu)$ is a $N(m, s^2)$. We conclude therefore that the conditional posterior distribution for $\mu$ is distributed as a $N\left[\left(\frac{m\frac{\sigma^2}{s^2} + \log \frac{S_t}{S_0} + \frac{t\sigma^2}{2}}{t + \frac{\sigma^2}{s^2}}\right), \frac{\sigma^2}{t + \frac{\sigma^2}{s^2}}\right]$, which is the same conditional posterior distribution as in Polson and Roberts (1994). □

## B.1 Consistency of posterior distributions for $\mu$ and $\sigma^2$

We now prove consistency of the posterior distibutions for both $\mu$ and $\sigma^2$ in the sense of definition 1. The posterior variance of $\mu$ is $\frac{\sigma^2}{t + \frac{\sigma^2}{s^2}}$ and goes to 0 as $t \to \infty$, which together with Chebychev's inequality shows consistency.

The posterior variance of $\sigma^2$ given the data and $\mu$ is given by Corollary 1.1 from Silva et al (2006):

$$Var(\sigma^2 \mid \lambda, \delta, \gamma) = \left(\frac{\delta}{\gamma}\right)^2 \left[\frac{K_{\lambda+2}(\gamma\lambda)}{K_\lambda(\gamma\lambda)} - \left(\frac{K_{\lambda+1}(\gamma\lambda)}{K_\lambda(\gamma\lambda)}\right)^2\right]$$

where $K_\nu(\omega)$ is the modified Bessel function of the third kind and equal to $K_\nu(\omega) = \frac{1}{2}\int_0^\infty x^{\nu-1}\exp\left\{-\frac{1}{2}\omega\left(x+x^{-1}\right)\right\}dx$. By noticing that

$$\lim_{\omega \to \infty} K_\nu(\omega) = 0$$

, $K_\nu(\omega) \approx \omega^{-\frac{1}{2}}\exp(-\omega)\sqrt{\frac{\pi}{2}}$ (for $\omega$ very big), and that $K_\nu(\omega) \geq 0$, we get the following inequalities for the posterior variance as $\omega \to \infty$:

$$\begin{aligned}
Var(\sigma^2 \mid \lambda, \delta, \gamma) &= \left(\frac{\delta}{\gamma}\right)^2 \left[\frac{K_{\lambda+2}(\gamma\lambda)}{K_\lambda(\gamma\lambda)} - \left(\frac{K_{\lambda+1}(\gamma\lambda)}{K_\lambda(\gamma\lambda)}\right)^2\right] \\
&\approx \left(\frac{\delta}{\gamma}\right)^2 \left[\frac{\omega^{-\frac{1}{2}}\exp(-\omega)\sqrt{\frac{\pi}{2}}}{\omega^{-\frac{1}{2}}\exp(-\omega)\sqrt{\frac{\pi}{2}}} - \left(\frac{\omega^{-\frac{1}{2}}\exp(-\omega)\sqrt{\frac{\pi}{2}}}{\omega^{-\frac{1}{2}}\exp(-\omega)\sqrt{\frac{\pi}{2}}}\right)^2\right] \\
&= 0 \text{ as } \omega \to \infty.
\end{aligned}$$

This last result shows that the posterior variance of $\sigma^2$ goes to zero as $\omega \to \infty$. Combining this last fact together with Chebychev's inequality shows consistency of the posterior probability distribution of both $\sigma^2$ and $\mu$.

## B.2 Application of lemmas 1 and 2 for Gibbs sampling

Given the full conditional distributions for both $\mu$ and $\sigma^2$ from lemma 1 and 2, we can use them to construct a Gibbs sampler in order to get the posterior probability distributions for both $\mu$ and $\sigma^2$ in the Black & Scholes model. We now present the Gibbs algorithm:

- Initialize both $\mu^0$ and $\sigma^{2,0}$ with starting values

- Draw $\mu^i \sim N\left[\left(\frac{m\frac{\sigma^{2,i-1}}{s^2} + \log\frac{S_t}{S_0} + \frac{t\sigma^{2,i-1}}{2}}{t + \frac{\sigma^{2,i-1}}{s^2}}\right), \frac{\sigma^{2,i-1}}{t + \frac{\sigma^{2,i-1}}{s^2}}\right]$

- Draw $\sigma^{2,i} \sim GIG(\lambda', \delta', \gamma')$, where the parameters $\lambda'$, $\delta'$, and $\gamma'$ are given by the above lemmas

Repeat for $i = 1, \cdots, I$ where $I$ is big enough.
The posterior for $\mu$ given $\sigma^2$ converges to its posterior mean since the posterior variance for both $\mu$ and $\sigma^2$ goes to 0 as $t \to \infty$ from subsection B.1.